\documentclass[aps,prd,twocolumn,floats,floatfix,showpacs,superscriptaddress,nofootinbib]{revtex4-1}

\usepackage{graphicx,amssymb,amsmath,amsthm,amsfonts,epsfig,epsf,fixmath}
\usepackage[usenames]{color}
\usepackage{epstopdf}

\usepackage{aas_macros}
\usepackage{bm}
\usepackage{dcolumn}
\usepackage[dvipsnames]{xcolor}
\usepackage[colorlinks]{hyperref}
\usepackage{cleveref}
\usepackage[utf8]{inputenc}
\usepackage{resizegather}

\begin{document}

\title{Hidden symmetry between rotational tidal Love numbers of spinning neutron stars}

\author{Gon\c{c}alo Castro}
	\email{goncalo.castro@uniroma1.it}
	\affiliation{Dipartimento di Fisica, ``Sapienza" Università di Roma \& Sezione INFN Roma1, Piazzale Aldo Moro 
5, 00185, Roma, Italy}
\author{Leonardo Gualtieri}
	\email{leonardo.gualtieri@roma1.infn.it}
	\affiliation{Dipartimento di Fisica, ``Sapienza" Università di Roma \& Sezione INFN Roma1, Piazzale Aldo Moro 
5, 00185, Roma, Italy}
\author{Paolo Pani}
	\email{paolo.pani@uniroma1.it}
	\affiliation{Dipartimento di Fisica, ``Sapienza" Università di Roma \& Sezione INFN Roma1, Piazzale Aldo Moro 
5, 00185, Roma, Italy}

\begin{abstract}
The coupling between the angular momentum of a compact object and an external tidal field gives rise to the
``rotational'' tidal Love numbers, which affect the tidal deformability of a spinning self-gravitating body and enter
the gravitational waveform of a binary inspiral at high post-Newtonian order.  We provide numerical evidence for a
surprising ``hidden'' symmetry among the rotational tidal Love numbers with opposite parities, which are associated to
perturbations belonging to separate sectors. This symmetry, whose existence had been suggested on the basis of a
Lagrangian description of the tidal interaction in a binary system, holds independently of the equation of state of the
star.
\end{abstract}

\maketitle

\section{Introduction and summary} \label{sec:intro}
When immersed in an external tidal field, a self-gravitating object gets deformed. The ``susceptibility'' to a tidal
deformation is measured by the so-called tidal Love numbers~(TLNs)~\cite{Murraybook,PoissonWill}, which depend on the
internal structure of the deformed body. The TLNs play a crucial role in gravitational-wave (GW) astronomy, most notably
to: i) constrain the equation of state~(EoS) of neutron stars~(NSs) through GW measurements of the tidal deformability
in the last stages of the
inspiral~\cite{Baiotti:2010xh,Baiotti:2011am,Vines:2011ud,Pannarale:2011pk,Vines:2010ca,Lackey:2011vz,Lackey:2013axa,
  Favata:2013rwa,Yagi:2013baa,Maselli:2013mva,Maselli:2013rza,DelPozzo:2013ala,TheLIGOScientific:2017qsa,Bauswein:2017vtn,
  Most:2018hfd,Harry:2018hke,Annala:2017llu,EoS-GW170817,Akcay:2018yyh} (see
Refs.~\cite{GuerraChaves:2019foa,Chatziioannou:2020pqz} for some recent reviews);~ii) constrain alternative theories of
gravity in an EoS-independent fashion~\cite{Yagi:2013awa,Yagi:2013bca} (see Ref.~\cite{Yagi:2016bkt} for a review);~iii)
test the nature of black holes with GW observations~\cite{Cardoso:2017cfl,Pani:2019cyc} (see Ref.~\cite{Cardoso:2019rvt}
for a review).

Clearly, the recent detections of coalescing NSs by LIGO/Virgo~\cite{TheLIGOScientific:2017qsa,Abbott:2020uma} give
strong motivation for further developments on this topic.  In particular, several binary NSs and mixed black 
hole-NS binaries will be detected in the future
LIGO/Virgo observation runs, possibly with higher signal-to-noise ratio than GW170817. While this will allow to put 
better constraints on the NS TLNs (and hence on the NS EoS), it makes it also urgent to develop waveform models that can
accurately take into account all possible effects related to the tidal deformability of
NSs~\cite{Abdelsalhin:2018reg,Jimenez-Forteza:2018buh,Banihashemi:2018xfb,Dietrich:2019kaq,Dietrich:2020eud,
  Henry:2020ski}.

Surprisingly, more than ten years after the seminal work by Flanagan and
Hinderer~\cite{Flanagan:2007ix,Hinderer:2007mb}, some properties of the TLNs are still being discovered and are still
not totally understood, in particular for what concerns the magnetic\,\footnote{The TLNs can be divided into two
  categories: \emph{electric} (or even parity), which are related to the mass multipole moments induced by the tidal
  field; and \emph{magnetic} (or odd parity), which are related to the induced current multipole moments and do not have
  an analog in Newtonian theory. See below for a formal definition.}
TLNs~\cite{Pani:2018inf,Poisson:2020mdi}. Furthermore, it was recently realized that the magnetic TLNs depend also on
the assumptions on the dynamics of the fluid within the star, namely whether the fluid is irrotational or static (see 
Sec.~\ref{subsec:statirr} for explicit definitions), with the former assumption being more physically
sound~\cite{Landry:2015cva,Landry:2015snx,Pani:2018inf}\,\footnote{We shall call static (irrotational) Love numbers
  those associated with a static (irrotational) fluid.}.
In this context, most of previous work on the tidal deformability had focused on nonspinning objects.
In recent years, there has been remarkable progress in extending the analysis to spinning compact objects.
The coupling between the object's angular momentum and the external tidal field introduces new families of so-called ``rotational'' 
TLNs~(RTLNs)~\cite{Pani:2015nua,Landry:2015cva,Landry:2015zfa,Landry:2017piv,Gagnon-Bischoff:2017tnz}.
Tidal deformations of slowly-spinning black holes were studied in Refs.~\cite{Poisson:2014gka,Pani:2015hfa}, which found
that the (R)TLNs of a black hole are zero~\cite{Binnington:2009bb,Damour:2009vw,Damour:2009va,Gurlebeck:2015xpa,Porto:2016zng,Hui:2020xxx,Charalambous:2021kcz} also in the
spinning case, at least to quadratic order in the spin in the axisymmetric case~\cite{Pani:2015hfa,Pani:2015nua} (see
also~\cite{Landry:2015cva,Landry:2015zfa}). Recently, using analytical-continuation
methods, it has been shown~\cite{LeTiec:2020spy} that the tidal field affects the non-axisymmetric multipole moments of a spining BH, already to linear order in the spin. As remarked in~\cite{Chia:2020yla}, the
non-vanishing quantities found in~\cite{LeTiec:2020spy} are associated with dissipative interactions, usually referred
to as tidal heating~\cite{Hartle:1973zz,PhysRevD.64.064004}. As argued in~\cite{Chia:2020yla}, the non dissipative (R)TLNs of a
Kerr black hole are identically zero, providing an ideal baseline for tests of the Kerr hypothesis with
GWs~\cite{Cardoso:2017cfl}. More generally, recent results~\cite{Poisson:2020mdi,Poisson:2020vap} show that, when the
tidal field depends on time or is not axisymmetric, the general picture of TLNs is more complicated than previously
expected. In this article we shall only consider a stationary, axisymmetric tidally deformed star, leaving the more
general case to a future analysis~\cite{inprep}.

Computing the RTLNs is rather involved, since it requires to work out the linear (gravitational and fluid) perturbations
of a spinning compact object and to solve the corresponding coupled system numerically.  Thus, it might not be
surprising that preliminary numerical computations of the RTLNs by different groups did not agree with each other.  In
particular, the analysis in Ref.~\cite{Gagnon-Bischoff:2017tnz} found disagreement with the RTLNs previously computed by
some of us~\cite{Pani:2015nua} (hereafter, Paper~I), especially for low-compactness NSs. We have found that the source of
disagreement is twofold. First, we have found an error in the numerical implementation of the equations in
Paper~I (now corrected in the computation presented below).
Second, the authors of~\cite{Gagnon-Bischoff:2017tnz} studied the irrotational RTLNs, arguing that in some cases they
coincide with the static RTLNs studied in Paper~I.  However, in general the irrotational RTLNs cannot be computed under
the assumption of stationarity. Properly including (slowly-varying) tidal perturbations allows to resolve the ambiguity
found in Ref.~\cite{Gagnon-Bischoff:2017tnz} and gives different irrotational RTLNs that do not coincide with the static
ones (although the differences are smaller than $5\%$). This point will be discussed in detail in a separate
publication~\cite{inprep}\,\footnote{By correcting the numerical implementation of Paper~I and integrating the field
  equations obtained with the assumptions of~\cite{Gagnon-Bischoff:2017tnz}, we find perfect agreement between the two
  approaches, up to numerical errors.}, while in this article we focus on the {\it static} RTLNs.

Although NS coalescing binaries are expected to have irrotational perturbations, static perturbations are useful to
elucidate a surprising feature of the RTLNs which we unveil in this work. Paper~I introduced \emph{four} independent
RTLNs to fully characterize the (quadrupolar and octupolar) tidal deformability of a spinning NS to linear order in the
spin and in the axisymmetric case, while the effective-field-theory Lagrangian developed in
Ref.~\cite{Abdelsalhin:2018reg} (hereafter, Paper~II) contains only \emph{two} parameters that govern the coupling
between the (quadrupolar and octupolar) tidal deformations of the body, its spin and the external tidal
field.  Thus, as argued in Paper~II, the Lagrangian approach seems to predict that the four RTLNs are not independent:
they should be related by two algebraic relations, and in fact two of them are simply proportional to the other two.

From a Lagrangian point of view it is natural to expect that opposite sectors are
coupled to each other. Indeed, a single interaction term in the schematic form
\begin{equation}
 {\cal L}({\cal A},\partial{\cal A},{\cal P},\partial{\cal P}) \supset  \alpha {\cal A} {\cal P}\,, \label{LagrangianCoupling}
\end{equation}
in the Lagrangian ${\cal L}$ gives rise --~using Euler-Lagrange equations~-- to related
coupling terms in the field equations for ${\cal A}$ and ${\cal P}$ which are both proportional to the single coupling constant
$\alpha$.

It is natural to ask whether similar relations are satisfied by the RTLNs, which are computed by solving the perturbation
equations of a single NS perturbed by a generic tidal source. Our analysis shows that this is indeed the case: we
computed static RTLNs (as we discuss in this paper, the Lagrangian constructed in Paper~II describes only {\it static}
perturbations) with different parities for various choices of the EoS and of the compactness, finding that the algebraic
relations derived in Paper~II are always satisfied, within the numerical errors, for low compactness and for any EoS (see
Fig.~\ref{fig:ratios}). When the compactness is large, there is still an EoS-independent (within numerical errors) relation between the RTLNs, but it deviates from the theoretical value predicted in Paper~II by up to~$6\%$.

These relations --~which are exact for low compactness and any EoS~-- imply the
existence of a ``\emph{hidden symmetry}'' among perturbations with even and odd parities. 
We use the term ``hidden symmetry'', which is stronger than ``universal relation'' (see~\cite{Yagi:2016bkt} for a
review) because, besides being EoS-independent within numerical uncertaintys, this symmetry is theoretically predicted by
a Lagrangian post-Newtonian (PN) approach.
Nonetheless, we stress that this hidden symmetry is truly unexpected and nontrivial from a perturbation-theory point of
view: the RTLNs that turn out to be proportional to each other belong to opposite parity sectors, so there is {\it a priori}
no reason why they should be related.
This is analogous to the symmetry between axial and polar
perturbations of a Schwarzschild black hole found by Chandrasekhar~\cite{Chandrasekhar:1985kt}, with the major
difference that the perturbation equations of compact stars depend on the EoS, making an analytical interpretation much
more challenging than in the case of black holes.  We argue that this symmetry should also affect irrotational
perturbations, although in that case it is likely to appear in a more involved form, requiring a more detailed study in
order to be elucidated~\cite{inprep}. In the case of large compactness, the hidden symmetry is only approximate, but an accurate
universal relation is still present.

The rest of this paper is organized as follows. In Sec.~\ref{sec:previous} we review Paper~I --~where the RTLNs are
introduced and the procedure for their numerical computation in terms of perturbations of a stationary NS is
described~-- and Paper~II --~where the effective Lagrangian describing two tidally interacting NSs is discussed. We also
briefly discuss, in Sec.~\ref{subsec:statirr}, the difference between static and irrotational perturbations.  Then, in
Sec.~\ref{sec:results} we discuss the numerical computation of the (static) RTLNs, showing that the afrorementioned
hidden symmetry is satisfied for different choices of the EoSs and of the compactness. We conclude in
Sec.~\ref{sec:conclusions}, where some open issues are discussed. Appendix~\ref{app:RTLNs} gives the explicit
expressions of the coefficients appearing in the perturbation equations, and Appendix~\ref{app:conversion} gives the
conversion factors between different (R)TLNs used in the literature.

\subsection{Notation and conventions}\label{subsec:notation}

We denote the speed of light in vacuum by $c$ and set the gravitational constant $G=1$. 
We shall mostly use units such that $c=1$, unless explicitly stated.
Latin indices run over three-dimensional spatial coordinates and are contracted with the Euclidean flat metric
$\delta_{ij}$; the antisymmetric Levi-Civita symbol in the Euclidean space is denoted by $\epsilon^{ijk}$. Greek indices
run over four-dimensional, spacetime coordinates.

Following the notation in~\cite{Thorne:1980ru} (see also~\cite{poisson2014gravity}), we use capital letters in the
middle of the alphabet $L,K$, etc., as shorthand for sets of indices $a_1\dots a_l$, $b_1\dots b_k$, etc. Round $()$,
square $[]$, and angular $\langle\rangle$ brackets enclosing the indices indicate symmetrization, antisymmetrization, 
and trace-free symmetrization, respectively.  We call symmetric trace-free (STF) those tensors $T^{a_1\cdots a_l}$ that 
are
symmetric on all indices and whose contraction of any two indices vanishes. For a generic vector $u^a$ we define
$u^{ab\cdots c}=u^au^b\dots u^c$ and $u^2=u^au^a$.

Functions and tensor fields on the two-sphere can be expanded in terms of tensor spherical harmonics: the scalar
spherical harmonics $Y^{l m}(\theta,\varphi)$, the vector spherical harmonics with even and odd parity, $(Y^{l
  m}_{,\theta},Y^{l m}_{,\varphi})$ and $(S^{l m}_\theta,S^{l m}_\varphi)=(-Y^{l
  m}_{,\varphi}/\sin\theta,\sin\theta Y^{l m}_{,\theta})$ respectively, etc. They can also be expanded in terms of
the STF tensors $n^L=n^{a_1}\cdots n^{a_l}$, where $n^a=(\sin\theta\cos\phi,\sin\theta\sin\phi,\cos\theta)$. Indeed,
$Y^{l m}(\theta,\phi)={\cal Y}^{l m}_{a_1\cdots a_l}n^{a_1\cdots a_l}$, where ${\cal Y}^{l m}_{a_1\cdots a_l}$
are constant coefficients.  Thus, any function $f(\theta,\phi)$ can be expanded as
\begin{equation}
f(\theta,\varphi)=\sum_{lm}f^{lm}Y^{lm}(\theta,\varphi)=\sum_lf_{a_1\cdots a_l}n^{a_1\cdots 
a_l}(\theta,\varphi)\,,\label{eq:harmstf}
\end{equation}
therefore $f_{a_1\cdots a_l}=\sum_mf^{lm}{\cal Y}^{l m}_{a_1\cdots a_l}$.

We denote the Geroch-Hansen multipole moments~\cite{Geroch:1970cd,Hansen:1974zz} by $M^L$ (mass) and $J^L$ (current), 
and
the Thorne multipole moments~\cite{Thorne:1980ru} by $Q^L$ (mass) and $S^L$ (current), see
also~\cite{Cardoso:2016ryw}. They are related by~\cite{1983GReGr..15..737G}
\begin{align}
M^L&=(2l-1)!!Q^L\,,\nonumber\\
J^L&=\frac{2l}{l+1}(2l-1)!!S^L\,.
\end{align}
We remind that the Geroch-Hansen multipole moments are defined in a coordinate-independent way, as tensors at infinity
generated by a set of potentials, while the Thorne moments are defined in terms of the asymptotic behvior of the
spacetime metric in asymptotically Cartesian mass-centered coordinates. We shall mostly use the Geroch-Hansen 
definition.

When the spacetime is symmetric with respect to an axis $\hat k$, the multipole moments can be written as $M^{i_1\cdots
  i_l}=(2l-1)!!M_lk^{i_1\cdots i_l}$, $J^{i_1\cdots i_l}=(2l-1)!!J_lk^{i_1\cdots i_l}$, and thus
\begin{align}
  Q^{i_1\cdots i_l}&=M_lk^{i_1\cdots i_l}\nonumber\\
  S^{i_1\cdots i_l}&=\frac{l+1}{2l}J_lk^{i_1\cdots i_l}\label{eq:transfmult}\,.
\end{align}
In this case, under the further assumption of symmetry with respect to the equatorial plane, the nonvanishing mass
multipole moments have even $l$, and the nonvanishing current multipole moments have odd $l$. When the spacetime is 
axial and equatorial symmetric, the Geroch-Hansen multipole moments can be also computed using Ryan's 
approach~\cite{Ryan:1995wh}, in terms of the geodesic properties of the spacetime metric.

The mass of a body and its angular momentum coincide with their $l=0$ mass and $l=1$ current multipole moments,
$M=M^0=Q^0$, $J=\sqrt{J^i J^i}=\sqrt{S^iS^i}$, respectively. We also define the dimensionless spin parameter of the body
as $\chi=J/M^2$, and the compactness as $C=M/R$, where $R$ is the stellar radius, defined by the location in which the
pressure of the fluid inside the star vanishes. Derivatives with respect to $t$ and $r$ are denoted with an overdot and
a prime, respectively.
\section{Review of previous work} \label{sec:previous}
Here we summarize the results of Paper~I and Paper~II. Since the notations and formalisms of these two papers are
different, we need to describe them in some detail, in order to compare, in Sec.~\ref{sec:results}, the results of the
numerical computation of the RTLNs, performed in the framework of Paper~I, with the theoretical predictions of Paper~II.

We remark that RTLNs have also been introduced, with a different notation,
in~\cite{Landry:2015cva,Landry:2015snx,Landry:2015zfa,Poisson:2016wtv,Landry:2017piv,Gagnon-Bischoff:2017tnz}, which
focus on the irrotational RTLNs.  A complete treatment of the irrotational RTLNs requires including slowly-varying
perturbations and will be discussed in details in a forthcoming publication~\cite{inprep}.

\subsection{Pani et al. (Paper~I)}\label{sec:pani}
In Paper~I (see also Ref.~\cite{Pani:2015hfa}), tidal deformations of rotating compact stars are 
studied by
considering stationary perturbations of a stationary, rotating star up to linear order in the spin (i.e. 
neglecting ${\cal O}(\chi^2)$ terms).
The perturbed metric can be written as $g_{\mu\nu}=g^{(0)}_{\mu\nu}+\delta g_{\mu\nu}$, where $g^{(0)}_{\mu\nu}$ is 
the background, whereas $\delta g_{\mu\nu}$ is the tidal perturbation. 
The background is described by Hartle's metric~\cite{Hartle:1967he,Thorne:1984mz}):
\begin{align}
  ds^{(0)\,2}&=g^{(0)}_{\mu\nu}dx^\mu dx^\nu=-e^\nu dt^2+e^\lambda dr^2 \nonumber\\
  &-2\sin^2\theta\omega r^2dtd\varphi+r^2d\Omega^2\,,
\end{align}
where $x^\mu=(t,r,\theta,\varphi)$, $d\Omega^2=d\theta^2+\sin^2\theta d\varphi^2$, $e^\lambda=(1-2{\cal M}/r)^{-1}$, and
the (radial) metric functions satisfy the set of ordinary differential equations:
\begin{align}
{\cal M}'&=4\pi r^2 P\,,\\
  \nu'&=\frac{2{\cal M}+4\pi r^2P}{(r-2{\cal M})}\,,\\
    P'&=-\frac{P+\rho}{2}\nu'\,,\\
{\tilde\omega}''&=\frac{4\pi r(P+\rho)(r{\tilde\omega}'+4\tilde\omega)}{r-2{\cal M}}-\frac{4}{r}{\tilde\omega}'\,,
\end{align}    
where $\tilde\omega=\Omega-\omega$, $\Omega$ is the fluid angular velocity, $P(r)$ and $\rho(r)$ are the pressure
and energy density of the fluid, respectively. The background four-velocity of the fluid is
$u^{(0)\mu}=e^{-\nu/2}(1,0,0,\Omega)$ and its stress-energy tensor is $T^{(0)\mu\nu}=(\rho+P)u^\mu
u^\nu+Pg^{(0)\mu\nu}$.  In vacuum, ${\cal M}(r)=M$ and $\omega(r)=2J/r^3$. 

The perturbations of the metric and of the fluid
four-velocity are expanded in tensor spherical harmonics, and are decomposed in even (or electric, or polar) and odd (or
magnetic, or axial) perturbations: $\delta g_{\mu\nu}=\delta g^{\rm (even)}_{\mu\nu}+\delta g^{\rm (odd)}_{\mu\nu}$,
with
(in the Regge-Wheeler gauge~\cite{regge:1957td})
\begin{align}
  \delta g_{\mu\nu}^{\rm (even)}dx^\mu dx^\nu&=e^\nu H_0^{lm}Y^{lm} dt^2+2H_1^{lm}Y^{lm} dtdr\label{expansion_h}\\
  &+H_2^{lm}Y^{lm} dr^2\,,\nonumber\\
  \delta g_{\mu\nu}^{\rm (odd)}dx^\mu dx^\nu&=(h_0^{lm}dt+h^{lm}_1dr)(S_\theta^{lm} d\theta+S_\phi^{lm} 
  d\varphi)\,,\nonumber
\end{align}
and $u^\mu=u^{(0)\mu}+\delta u^\mu$. In Paper~I the perturbations were assumed to be {\it static} (see
Sec.~\ref{subsec:statirr}), i.e. $\delta g_{\mu\nu,0}=0$ and $\delta u^i=0$ ($i=1,2,3$).  Thus, the perturbations with
even parity are described by the functions $(H_0(r)^{lm}(r),H_1^{lm}(r),H_2^{lm}(r),K^{lm}(r))$, and those with odd
parity are described by the functions $(h_0^{lm}(r),h_1^{lm}(r))$. It was also assumed that the perturbations are
axisymmetric, and thus they have $m=0$ (with symmetry axis parallel to the body's angular momentum); we remark that
$m\neq0$ tidal perturbations of a spinning object would induce precession and hence a weak time-dependence of the
perturbed system~\cite{Thorne:1997kt}. Thus, assuming static perturbations implies $m=0$.

The field equations at first order in the spin mix the perturbations having a given (polar or axial) parity and harmonic
index $l$ with those having opposite parity and harmonic index $l\pm1$ (see Ref.~\cite{Pani:2013pert} for a
review). Thus, it is possible to define ``polar-led'' and ``axial-led'' perturbations; the former are induced by a
purely electric tidal field, the latter by a purely magnetic tidal field. The polar-led system has the form (leaving
implicit the index $m=0$)
\begin{align}
{\cal D}^{{\rm pol}\,(l)}[H_0^{l}]&=0\,,\nonumber\\
{\cal D}^{{\rm ax}\,(l+1)}[h_0^{l+1}]&=S_+^{{\rm pol}\,(l)}[H_0^{l}]\,,\nonumber\\
{\cal D}^{{\rm ax}\,(l-1)}[h_0^{l-1}]&=S_-^{{\rm pol}\,(l)}[H_0^{l}]\,
\label{eq:polarled}
\end{align}
where
\begin{align}
  {\cal D}^{{\rm pol}\,(l)} H_0^l &= \frac{d^2 H_0^l}{dr^2} + C_1^{{\rm pol}\,(l)}(r) \frac{d H_0^l}{dr} +
  C_0^{{\rm pol}\,(l)}(r)H_0^l 
	\label{eq:dpol}\\
	      {\cal D}^{{\rm ax}\,(l)} h_0^l& = \frac{d^2 h_0^l}{dr^2} + C_1^{{\rm ax}\,(l)}(r) \frac{d h_0^l}{dr} +
              C_0^{{\rm ax}\,(l)}(r)h_0^l\,.
	\label{eq:dax}
\end{align}
The perturbation $H_0^{l}$ is at zero order in the spin, while the perturbations $h_0^{l\pm1}$ are at first order in the
spin, and vanish in the $\Omega\to0$ limit. The other perturbation functions can be obtained from $H_0^{l}$ and
$h_0^{l\pm1}$ through algebraic relations. Similarly, the axial-led system has the form
\begin{align}
{\cal D}^{{\rm ax}\,(l)}[h_0^{l}]&=0\,,\nonumber\\
{\cal D}^{{\rm pol}\,(l+1)}[H_0^{l+1}]&= S_+^{{\rm ax}\,(l)}[h_0^{l}]\,,\nonumber\\
{\cal D}^{{\rm pol}\,(l-1)}[H_0^{l-1}]&= S_-^{{\rm ax}\,(l)}[h_0^{l}]\,.\label{eq:axialled}
\end{align}
In this case $h_0^{l}$ is at zero order in the spin, while the perturbations $H_0^{l\pm1}$ are at first order in the
spin, and vanish in the $\Omega\to0$ limit. The explicit forms of the coefficients $C_{0,1}^{{\rm pol}\,(l)}$,
$C_{0,1}^{{\rm ax}\,(l)}$ and of the sources $S^{{\rm pol}\,(l)}_\pm$, $S^{{\rm ax}\,(l)}_\pm$ is given in
Appendix~\ref{app:RTLNs}.  The other perturbation functions can be obtained from $h_0^{l}$ and $H_0^{l\pm1}$ through
algebraic relations.

The source of the perturbations is an asymptotic tidal field, described by the electric and magnetic tidal tensors,
${\cal E}^{(l)}_m$ and ${\cal B}^{(l)}_m$, respectively. The leading-order asymptotic expansion of the metric (as $r\gg
M$) in terms of these tensors reads:
\begin{align}
g_{tt}&\to-\sum_{l\ge2,m}\frac{2}{l(l-1)}{\cal E}^{(l)}_mY^{lm}(\theta)r^l\nonumber\\
g_{t\varphi}&\to\sum_{l\ge2}\frac{2}{3l(l-1)}{\cal B}^{(l)}_mS_\varphi^{lm}(\theta)r^{l+1}\,.\label{eq:defcEcB}
\end{align}
In the case of an axisymmetric perturbation, only the $m=0$ tidal fields, ${\cal E}^{(l)}_0$, ${\cal B}^{(l)}_0$,
contribute.  Note that the metric~\eqref{eq:defcEcB} is not asymptotically flat. Indeed, it only describes the
spacetime at $r\lesssim r_{\rm ts}$ where $r_{\rm ts}$ is the location of the generic source of the tidal field.

As a result of the tidal field, the mass and current multipole moments are deformed; in linear perturbation theory,
these deformations are proportional to the tidal fields themselves.  At zero-th order in the spin, the electric
(magnetic) tidal field affects the mass (current) multipole moment with the same value of $l$. The proportionality
constants
\begin{align}
  \lambda^{(l)}_E&\equiv\frac{\partial M_l}{\partial{\cal E}^{(l)}_{0}}\,,\nonumber\\
  \lambda^{(l)}_M&\equiv\frac{\partial J_l}{\partial{\cal B}^{(l)}_{0}}\,,\label{eq:defTLN}
\end{align}
are the {\it relativistic TLNs}~\cite{Damour:2009vw,Binnington:2009bb}.  At first order in the spin, the tidal field
with a given parity and harmonic index $l$ affects the tidal field with opposite parity and harmonic index $l\pm1$. The
proportionality constants
\begin{align}
\lambda_E^{(ll')}&=\frac{\partial M_l}{\partial{\cal B}^{(l')}_{0}}\,,\nonumber\\
\lambda_M^{(ll')}&=\frac{\partial J_l}{\partial{\cal E}^{(l')}_{0}}\,,\label{eq:defRTLN}
\end{align}
with $l'=l\pm1$, are called {\it relativistic RTLNs}, see e.g. Paper~I 
and~\cite{Pani:2015hfa,Landry:2015cva,Landry:2015zfa}.

Since $[{\cal E}^{(l)}_0]=[{\cal B}^{(l)}_0]=({\rm mass})^{-l}$, 
$[M_l]=[J_l]=({\rm mass})^{l+1}$, and the RTLNs are
proportional to the dimensionless spin, the {\it dimensionless} TLNs and RTLNs can be defined as
\begin{align}
\tilde{\lambda}^{(l)}_{E/M}&\equiv\frac{\lambda^{(l)}_{E/M}}{M^{2l+1}}\,,\nonumber\\
\tilde{\lambda}^{(ll')}_{E/M}&\equiv\frac{\lambda^{(ll')}_{E/M}}{\chi M^{l+l'+1}}\,.
\end{align}
Note that the dimensionless RTLNs defined above are also independent of the spin.
In terms of these quantities, the axisymmetric deformations of the quadrupole and octupole
moments, to linear order in the tidal tensor and to linear order in the spin, are:
\begin{align}
  \frac{M_2}{M^3}&=\tilde{\lambda}^{(2)}_E \tilde{\cal E}^{(2)}_0+\chi\tilde\lambda^{(23)}_E\tilde{\cal B}_0^{(3)}\nonumber\\
  \frac{M_3}{M^4}&=\tilde{\lambda}^{(3)}_E\tilde{\cal E}^{(3)}_0+\chi\tilde\lambda^{(32)}_E\tilde{\cal B}_0^{(2)}\nonumber\\
  \frac{J_2}{M^3}&=\tilde{\lambda}^{(2)}_M\tilde{\cal B}^{(2)}_0+\chi\tilde\lambda^{(23)}_M\tilde{\cal E}_0^{(3)}\nonumber\\
  \frac{J_3}{M^4}&=\tilde{\lambda}^{(3)}_M\tilde{\cal B}^{(3)}_0+\chi\tilde\lambda^{(32)}_M\tilde{\cal E}_0^{(2)}\,,
  \label{eq:adiabrelp}
\end{align}
where we defined the dimensionless tidal tensors $\tilde {\cal E}^{(l)}_m ={\cal E}^{(l)}_m M^l$ and $\tilde {\cal
  B}^{(l)}_m ={\cal B}^{(l)}_m M^l$.  Note that if the system is symmetric  with respect to the equatorial plane, $M_3=J_2={\cal
  E}_0^{(3)}={\cal B}_0^{(2)}=0$.

The Love numbers can be computed by solving the systems in Eqs.~\eqref{eq:polarled} and \eqref{eq:axialled} for $l=2,3$
with the tidal sources
${\cal E}^{(2)}_0$, ${\cal E}^{(3)}_0$, ${\cal B}^{(2)}_0$, ${\cal B}^{(3)}_0$. The analytic solution oustide the star
has been explicitly derived in Ref.~\cite{Pani:2015hfa}, in terms of a set of integration constants. Solving the equations
inside the star, with the assumptions of regularity at the center and smooth boundary conditions at the surface $r=R$ of
the star, fixes the integration constants, and thus gives the explicit value of the TLNs and of the RTLNs, which depend
on the EoS of the star. 

\subsection{Abdelsalhin et al. (Paper~II)}\label{sec:abdelsalhin}
In Paper~II (see also Ref.~\cite{Abdelsalhin:2019ryu}), the leading-order contribution of the RTLNs to the PN waveform
of coalescing compact binaries was computed, together with the spin corrections to the tidal deformability terms. These
terms appear at $6.5$PN order (i.e., they are suppressed by a factor $v^{13}$, where $v$ is the orbital velocity of the
binary) relative to the leading-order contribution to the GW phase (and by a factor $v^3$ relative to the 
leading-order TLN term entering at $5$PN order); they are obtained by generalizing previous
results~\cite{Vines:2010ca,Vines:2011ud}, where the $6$PN tidal term in the GW phase of nonrotating compact coalescing
binaries was derived.

The motion of the binary is described in terms of a Lagrangian function
\begin{equation}
{\cal L}={\cal L}(z^i,{\dot z}^i,{\ddot z}^i,M_A,J_A,Q^L_A,{\dot Q}^L_A,S^L_A,{\dot S}^L_A)\,.
\end{equation}
Here $z^i=z^i_1-z_2^i$ is the relative position of the binary in the harmonic, conformally Cartesian coordinate frame
(defined everywhere outside the strong-field region of the compact bodies) in which the PN approximation is defined; an
overdot denotes a derivative with respect to the coordinate time in this frame; the index $A=1,2$ refers to the two
bodies in the binary\,\footnote{Note that units used in Paper~II are such that $G=1$, while the speed of light $c$ is 
retained as a dimensionful quantity in order to keep track of the different PN orders (terms of the $n$-th PN order are
$O(c^{-2n})$).}. 

Each body is characterized by its mass $M_A$, by its angular momentum $J_A$ and by the higher-order multipole moments
$Q_A^L$, $S_A^L$ (with $l\ge2$), using Thorne's definition (see Sec.~\ref{subsec:notation}), which are induced by the
tidal field of the companion. The multipole expansion is truncated to the octupole $(l=3)$, and the rotation is included
at first order in the spin. The next-to-leading order quadrupolar contributions are included, whereas the octupolar
contributions are truncated at the leading order. This truncation is sufficient to determine the tidal waveform up to
$6.5$PN order. Moreover, for simplicity the quadrupole and octupole moments of body~$1$ are set to zero, thus including
only the multipole moments induced from body~$1$ to body~$2$; the moments induced from body~$2$ to body~$1$ can be
obtained {\it a posteriori} with a simple exchange of indices. Thus, the quadrupole and octupole moments are simply
denoted as $Q^L\equiv Q_2^L$ and $S^L\equiv S_2^L$, respectively.

The Lagrangian can be written as the sum of an orbital Lagrangian, which depends on the orbital motion and on the
multipole moments, and an internal Lagrangian, which depends on the internal degrees of freedom of body~$2$ only:
\begin{align}
  {\cal L} &= {\cal L}_{\rm orb}(z^i,{\dot z}^i,{\ddot z}^i,M_A,J_A,Q^{ab},{\dot 
Q}^{ab},Q^{abc},S^{ab},S^{abc})\nonumber\\
  &+{\cal L}_2^{\rm int}(Q^{ab},Q^{abc},S^{ab},S^{abc})\,.
\label{eq:lagrangian}
\end{align}
We remark that the terms in ${\dot Q}^{ab}$ in ${\cal L}_{\rm orb}$ contribute to the next-to-leading order corrections
to the gravitational waveform. The time derivatives of the other moments are subleading.

The expression of the orbital Lagrangian is uniquely determined by imposing that its variation with respect to the
coordinate separation leads to the PN orbital equations of motion. The variation of the orbital Lagrangian with respect
to the multipole moments gives the electric and magnetic tidal tensors, which are denoted by $G^L$ and $H^L$, 
respectively: 
\begin{align}
G^{ab}&=\frac{\partial{\cal L}_{\rm orb}}{\partial Q^{ab}}  -\frac{d}{dt}\frac{\partial{\cal L}_{\rm orb}}{\partial 
{\dot Q}^{ab}}
\nonumber\\
G^{abc}&=\frac{\partial{\cal L}_{\rm orb}}{\partial Q^{abc}}\nonumber\\
H^{ab}&=\frac{\partial{\cal L}_{\rm orb}}{\partial S^{ab}}\nonumber\\
H^{abc}&=\frac{\partial{\cal L}_{\rm orb}}{\partial S^{abc}}\,,\label{eq:deftidaltensors}
\end{align}  
and are defined, as in Paper~I, from the asymptotic behavior of the metric around each of the bodies
composing the system. For each body one can define a ``buffer region'', far enough from the body so that the 
gravitational field is weak, but close enough so that the effect of the other body appears as a tidal field. In the 
buffer region around the body~$A$, $x^i=z_A^i+y^i$, and the tidal contribution to the metric is (see Eqs.~(1.63), (1.76)
in Paper~II):
\begin{align}
g_{tt}&=\dots+\sum_{l\ge2,m}\frac{2}{l!}G^{lm}Y^{lm}(\theta)r^l\nonumber\\
g_{ti}&=\dots-\sum_{l\ge2,m}\frac{1}{(l+1)!}H^{lm}S_\varphi^{lm}(\theta)r^{l+1}\,,\label{eq:defcEcB2}
\end{align}
where $G^{lm}Y^{lm}=G^{a_1\cdots a_l}n^{a_1\cdots a_l}$, $H^{lm}Y^{lm}=G^{a_1\cdots
  a_l}n^{a_1\cdots a_l}$, see Eq.~\eqref{eq:harmstf}. Note that this is equivalent to Eq.~\eqref{eq:defcEcB} but with a 
different notation, see Appendix~\ref{app:conversion}. 

At first order in the perturbation, the multipole moments are linear in the tidal tensors which induce them; at linear
order in the spin, they can be written (keeping factors of $c$ for clarity)
\begin{align}
Q^{ab}&=\lambda_2 G^{ab}+\frac{\lambda_{23}}{c^2}J^cH^{abc}\,,\nonumber\\
Q^{abc}&=\lambda_3 G^{abc}+\frac{\lambda_{32}}{c^2}J^{<c} H^{ab>}\,,\nonumber\\
S^{ab}&=\frac{\sigma_2}{c^2} H^{ab}+\sigma_{23}J^cG^{abc}\,,\nonumber\\
S^{abc}&=\frac{\sigma_3}{c^2} H^{abc}+\sigma_{32}J^{<c} G^{ab>}\,,\label{eq:adiabatic}
\end{align}
where $\lambda_l$, $\sigma_l$ are the electric and magnetic TLNs, and $\lambda_{ll'}$, $\sigma_{ll'}$ are the electric
and magnetic RTLNs. They are defined with a different normalization with respect to those introduced in Paper~I and in
Sec.~\ref{sec:pani}; the conversion factors among them are (see Appendix~\ref{app:conversion})
\begin{align}
\lambda_2&=-\frac{4}{3}\sqrt{\frac{\pi}{5}}M^5{\tilde\lambda}^{(2)}_E\nonumber\\
\lambda_3&=-\frac{4}{5}\sqrt{\frac{\pi}{7}}M^7{\tilde\lambda}^{(3)}_E\nonumber\\
\sigma_2&=-\frac{1}{2}\sqrt{\frac{\pi}{5}}M^5{\tilde\lambda}^{(2)}_M\nonumber\\
\sigma_3&=-\frac{1}{5}\sqrt{\frac{\pi}{7}}M^7{\tilde\lambda}^{(3)}_M\nonumber\\
\lambda_{23}&=-\sqrt{\frac{\pi}{7}} M^4{\tilde\lambda}^{(23)}_E\nonumber\\
\lambda_{32}&=-\frac{3}{4}\sqrt{\frac{\pi}{5}} M^4{\tilde\lambda}^{(32)}_E\nonumber\\
\sigma_{23}&=-2\sqrt{\frac{\pi}{7}} M^4{\tilde\lambda}^{(23)}_M\nonumber\\
\sigma_{32}&=-16\sqrt{\frac{\pi}{5}} M^4{\tilde\lambda}^{(32)}_M\label{eq:transf}\,.
\end{align}

Equations~\eqref{eq:adiabatic} are called {\it adiabatic relations} because the TLNs and the RTLNs are assumed to be
constant, neglecting the oscillatory response to a variation of the tidal field; this adiabatic approximation is
violated in the final stages of the coalescence~\cite{Maselli:2012zq,Steinhoff:2016rfi}. Note that (at variance with
Paper~I) in Paper II the multipole moments and the tidal tensor can change with time; so, for instance, the orbital
Lagrangian (but not the internal Lagrangian, see Sec.~\ref{subsec:statirr}) depends on the multipole moments and on
their time derivatives. However, in the adiabatic approximation the time dependence of the tidal fields is neglected.
This implies that the nonaxisymmetric contribution of the tidal fields and of the multipole moments vanishes, since
stationary perturbations must be axisymmetric. In the general case (not discussed in this paper), the TLNs and the RTLNs
are {\it matrices} in the STF framework, which correspond, in the harmonic basis, to Love numbers depending on the
indices $l$ and $m$~\cite{LeTiec:2020spy}.

The internal Lagrangian ${\cal L}_2^{\rm int}(Q^{ab},Q^{abc},S^{ab},S^{abc})$ only depends on the internal degrees of
freedom of the body~$2$, and is determined by imposing that the variation of
${\cal L}={\cal L}_{\rm orb}+{\cal L}_2^{\rm int}$ with respect to the multipole moments yields the adiabatic
relations~\eqref{eq:adiabatic}. This gives
\begin{align}
{\cal L}_2^{\rm 
int}&=-\frac{1}{4\lambda_2}Q^{ab}Q^{ab}-\frac{1}{12\lambda_3}Q^{abc}Q^{abc}-\frac{1}{6\sigma_2}S^{ab}S^{ab}\nonumber\\
&-\frac{1}{16\sigma_3}S^{abc}S^{abc}+\alpha J_2^aQ^{bc}S^{abc}\nonumber\\
&+\beta J^a_2S^{bc}Q^{abc}\,,\label{Lintpaper2}
\end{align}
where $\alpha$ and $\beta$ are related to the RTLNs. Indeed, using Eqs.~\eqref{eq:deftidaltensors} the variation of the 
Lagrangian with respect to the multipole moments
gives Eq.~\eqref{eq:adiabatic}, with
\begin{align}
&\lambda_{23}=2\lambda_2\sigma_3\alpha~~~~~\lambda_{32}=6\lambda_3\sigma_2\beta\nonumber\\
&\sigma_{23}=3\lambda_3\sigma_2\beta~~~~~\sigma_{32}=8\lambda_2\sigma_3\alpha\,.\label{eq:identifications}
\end{align}
Remarkably, Eqs.~\eqref{eq:identifications} lead to
\begin{equation}
\sigma_{32}=4\lambda_{23}\,,~~~~~\sigma_{23}=\frac{1}{2}\lambda_{32}\,,\label{eq:prop}
\end{equation}
which, in the notation of Paper~I (see Eqs.\eqref{eq:transf}), gives 
\begin{eqnarray}
{\tilde\lambda}^{(32)}_M&=&\frac{9}{4}\sqrt{\frac{5}{7}}{\tilde\lambda}^{(23)}_E\,,
\label{eq:prop2}\\
\tilde{\lambda_M^{(23)}} &=& \frac{1}{3}\sqrt{\frac{7}{5}}\tilde{\lambda}_E^{(32)}\,. \label{eq:prop2b}
\end{eqnarray}
We stress that the above relations among the RTLNs follow from the use of the Lagrangian formulation, which is also
instrumental to obtain the gravitational waveform. Remarkably, the magnetic-led RTLN $\sigma_{32}$ (i.e.,
${\tilde\lambda}^{(32)}_M$) and the electric-led RTLN $\lambda_{23}$ (i.e., ${\tilde\lambda}^{(23)}_E$) are obtained from
equations involving perturbations with \emph{opposite} parities. Therefore, from the perturbation theory point of view,
there is {\it a priori} no reason to expect these pairs of RTLNs to be related. Nonetheless, in the next section we will
confirm that Eqs.~\eqref{eq:prop2} and \eqref{eq:prop2b} hold true by computing the RTLNs of a spinning NS.  As
discussed in the introduction, this relation reveals the existence of a new type of hidden symmetry in the structure of
compact stars: a universal relation satisfied for \emph{any} EoS, which is exact for small compactness and weakly
violated at large compactness.

\begin{figure*}[th]
\includegraphics[width=0.48\textwidth]{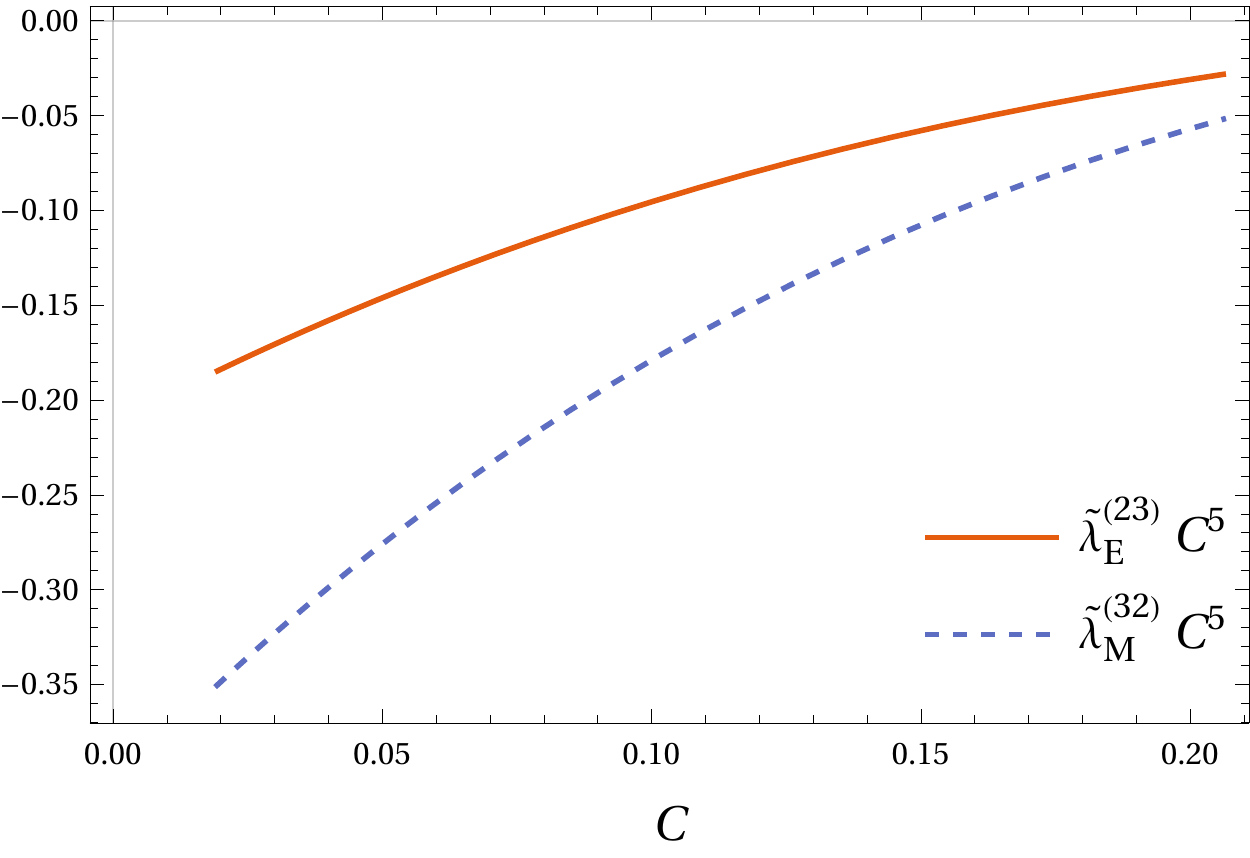}
\includegraphics[width=0.48\textwidth]{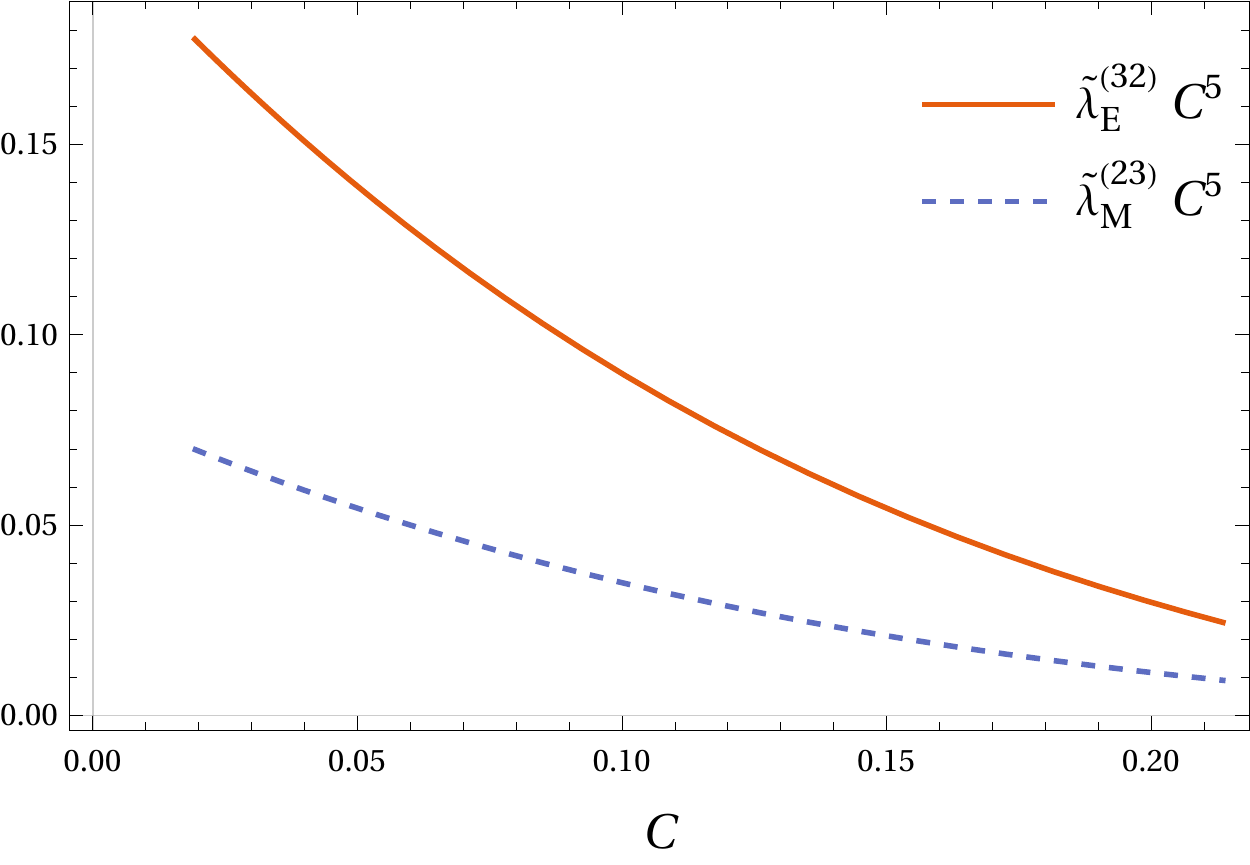}
\caption{Electric-led and magnetic-led RTLNs for a static fluid with polytropic 
	EoS with $n=1$. These results are obtained after correcting the numerical implementation of Paper~I.}
\label{1love}
\end{figure*}

\subsection{Static and irrotational relativistic TLNs}\label{subsec:statirr}
In Paper~I it was assumed that the fluid is {\it static}, i.e. $\delta u^i=0$. More recently, it was found that the
appropriate stationary limit of a time-dependent compact star has an irrotational fluid~\cite{Landry:2015cva}, in which
$\delta g_{\mu\nu,0}=0$ as for static perturbations, but $\delta u^i$ can have an azimuthal component. In the
irrotational case, the perturbation of the vorticity tensor,
\begin{equation}
  \omega_{\alpha\beta}=\nabla_{\alpha}(hu_\beta)-\nabla_\beta(hu_\alpha)\,,
\end{equation}
(with $h=(\rho+P)/n$, $n$ baryonic number density) identically vanishes.
Conversely the static fluid, although mathematically
consistent (it is an admissible solution of the field equations), cannot be retrieved as the static limit of a
time-dependent solution, and thus should not be considered as physically sound.

For a non-rotating NS the irrotationality condition simply reduces to the vanishing of the covariant velocity
perturbation, $\delta u_i=0$~\cite{Landry:2015cva,Pani:2018inf}, leading to $\delta u^\phi\propto \delta g_{t\phi}$.  As
discussed in Ref.~\cite{Pani:2018inf}, this choice corresponds to the magnetic TLNs computed by Damour and
Nagar~\cite{Damour:2009vw}. For a rotating star the irrotational condition is much more involved, it does not reduce to
a simple condition on the four-velocity components (see e.g.~\cite{Lockitch:2000aa}).

As noted in~\cite{Gupta:2020lnv}, this characterization of static and irrotational fluid configurations can be easily
rephrased in a Lagrangian framework. If the internal Lagrangian does not depend on time, its variations yield the static
perturbations. The irrotational perturbations, instead, are the zero-frequency limit of the equations obtained from a
time-dependent Lagrangian.

In Paper~II it was assumed that the Lagrangian ${\cal L}_{\rm int}$, describing the internal degrees of freedom of the
star, depends on the multipole moments but not on their time derivatives: ${\cal L}_{\rm int}={\cal L}_{\rm
  int}(Q^L_A,S^L_A)$, see Eq.~\eqref{Lintpaper2}. Therefore, the adiabatic equations~\eqref{eq:adiabatic}, arising from the
variations $\partial{\cal L}/\partial Q^L_A$ and $\partial{\cal L}/\partial S^L_A$, correspond to {\it static
  perturbations}. Thus, the hidden symmetry~\eqref{eq:prop2}-\eqref{eq:prop2b} refers to static perturbations as well.
In order to extend the results of Paper~II to {\it irrotational} perturbations, we should consider an internal
Lagrangian which depends (like the orbital Lagrangian) to ${\dot Q}^L_A,{\dot S}^L_A$ as well, compute the variations
with respect to the multipole moments, and finally consider the zero-frequency limit; we leave this computation for
future work~\cite{inprep}.
\begin{figure*}[th]
\includegraphics[width=0.47\textwidth]{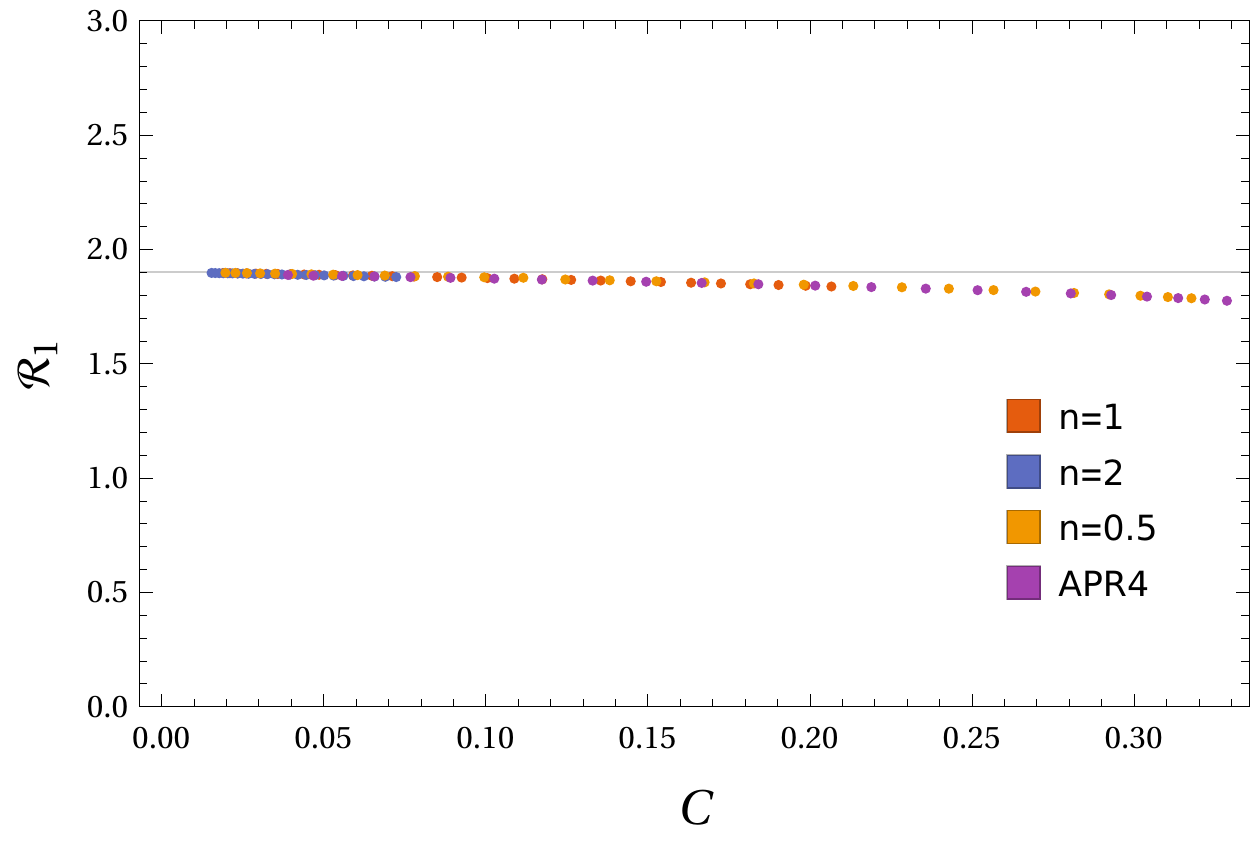}
\includegraphics[width=0.47\textwidth]{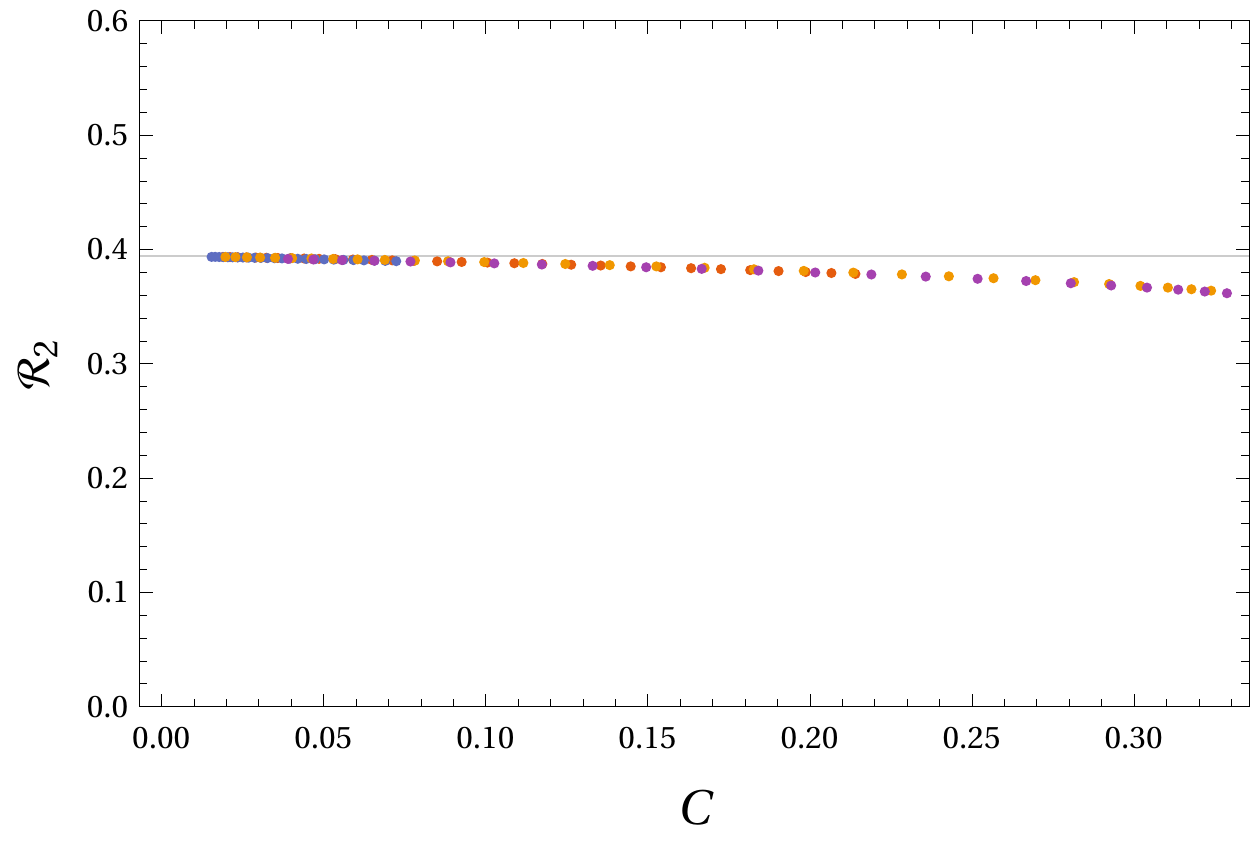}
\caption{Ratios of RTLNs ${\cal R}_1$ (left panel) and ${\cal R}_2$ (right panel) [see Eq.~\eqref{eq:prop2r}]
  as a function of the compactness $C$, for a polytropic EoS with $n=0.5$,
  $n=1$, $n=2$ and for the APR4 EoS. The horizontal line denotes the theoretical prediction
  from the Lagrangian formulation [Eq.~\eqref{eq:prop2r}].}
\label{fig:ratios}
\end{figure*}

\section{Hidden symmetry of the RTLNs} \label{sec:results}
We shall now explicitly compute the static RTLNs for different EoSs. We shall then verify whether the
relations~\eqref{eq:prop2}, \eqref{eq:prop2b} are satisfied.
\subsection{Computation of the static RTLNs}
In order to compute the magnetic RTLNs~\eqref{eq:defRTLN}
\begin{equation}
  \lambda_M^{(ll')}=\frac{\partial J_l}{\partial{\cal E}^{(l')}_{0}}
\end{equation}
with $l'=l\pm1$, and the rescaled quantities ${\tilde\lambda}_M^{(ll')}=\lambda_M^{(ll')}/(M^{l+l'+1}\chi)$, 
we have to determine the current multipole perturbations $J_l$
induced by an electric tidal field ${\cal E}^{(l')}_0$~\eqref{eq:defcEcB}; thus, we have to find the axial parity
perturbations $h^l_0$ induced by polar parity perturbations $H_0^{l'}$, by solving the polar-led
equations~\eqref{eq:polarled}. For ${\tilde\lambda}^{(23)}_M$, we have to consider the perturbations $h_0^2$ induced by
$H_0^3$, and Eqs.~\eqref{eq:polarled} reduce to
\begin{align}
  {\cal D}^{{\rm pol}\,(3)}[H_0^{3}]&=0\label{eq:3Estat}\\
{\cal D}^{{\rm ax}\,(2)}[h_0^{2}]&=S_+^{{\rm pol}\,(3)}[H_0^{3}]\,,\label{eq:23Mstat}
\end{align}
while for
${\tilde\lambda}^{(32)}_M$, we have to consider the perturbations $h_0^3$ induced by $H_0^2$, and Eqs.~\eqref{eq:polarled}
reduce to
\begin{align}
  {\cal D}^{{\rm pol}\,(2)}[H_0^{2}]&=0\label{eq:2Estat}\\
{\cal D}^{{\rm ax}\,(3)}[h_0^{3}]&=S_+^{{\rm pol}\,(2)}[H_0^{2}]\,.\label{eq:32Mstat}
\end{align}
Similarly, to compute the electric  RTLNs~\eqref{eq:defRTLN}
\begin{equation}
\lambda_E^{(ll')}=\frac{\partial M_l}{\partial{\cal B}^{(l')}_{0}}
\end{equation}
with $l'=l\pm1$,  and the rescaled quantities ${\tilde\lambda}_E^{(ll')}=\lambda_E^{(ll')}/(M^{l+l'+1}\chi)$, 
we have to determine the mass multipole perturbations $M_l$ induced by a magnetic
tidal field ${\cal B}^{(l')}_0$~\eqref{eq:defcEcB}.
Thus, we have to find the
polar parity perturbations $H^l_0$ induced by axial parity perturbations $h_0^{l'}$,
by solving the axial-led equation~\eqref{eq:axialled}. For ${\tilde\lambda}^{(23)}_E$, we have to consider the perturbations
$H_0^2$ induced by $h_0^3$, and Eqs.~\eqref{eq:axialled} reduce to
\begin{align}
  {\cal D}^{{\rm ax}\,(3)}[h_0^{3}]&=0\label{eq:3Mstat}\\
{\cal D}^{{\rm pol}\,(2)}[H_0^{2}]&=S_+^{{\rm ax}\,(3)}[h_0^{3}]\,,\label{eq:23Estat}
\end{align}
while for
${\\tildelambda}^{(32)}_E$, we have to consider the perturbations $H_0^3$ induced by $h_0^2$, and Eqs.~\eqref{eq:axialled}
reduce to
\begin{align}
  {\cal D}^{{\rm ax}\,(3)}[h_0^{2}]&=0\label{eq:2Mstat}\\
{\cal D}^{{\rm pol}\,(3)}[H_0^{3}]&=S_+^{{\rm ax}\,(2)}[h_0^{2}]\,.\label{eq:32Estat}
\end{align}
The explicit form of Eqs.~\eqref{eq:32Mstat}-\eqref{eq:32Estat} is given in Eqs.~\eqref{eq:dpol}, \eqref{eq:dax} and in
Appendix~\ref{app:RTLNs}.

For concreteness, let us consider a quadrupolar electric tidal perturbation, $\mathcal{E}_0^{(2)}$ (the computation of
other multipoles follows straightforwardly).  In practice, we start by solving Eq.~\eqref{eq:2Estat} and then use
its solution to source Eq.~\eqref{eq:32Mstat}. Outside the star, we can solve the entire system analytically, obtaining
a solution that satisfies the asymptotic behavior~\eqref{eq:defcEcB}.  The full treatment of the separation of the tidal
and response parts of the solutions, as well as the definition of the solutions' free constants, is the same as in
Paper~I.

The interior solutions have to be computed numerically. We start by performing an asymptotic expansion of
Eq.~\eqref{eq:2Estat} at $r=0$ to obtain the initial conditions (up to an overall constant) and integrate up to the
radius of the star.  We then match the two (interior and exterior) solutions and their first radial derivatives at the
radius $R$, obtaining values for the free constants of the exterior solution (specifically, $\gamma_2$ and $\alpha_2$ as
defined in Paper~I) with which we compute the ($l=2$, electric) TLN of the non-spinning star,
${\tilde\lambda}_E^{(2)}$. Solving Eq.~\eqref{eq:32Mstat} follows a similar procedure, with the difference that --~being
an inhomogeneous equation~-- its full solution is obtained as a linear combination of a particular solution and the
solution of the corresponding homogeneous equation. The arbitrary multiplicative factor in front of the homogeneous part
--~as well as the value of the free single constant of the exterior solution ($\gamma^*_{32}$ as defined in Paper~I)~--
can be obtained via the matching at the radius of the star $R$. After this procedure one can simply extract the
corresponding RTLN, ${\tilde\lambda}_M^{(32)}$.

We compute the electric and magnetic RTLNs with $l,l'=2,3$ and $l,l'=3,2$ for different values of the compactness
$C=M/R$, for a sample of EoS consisting in the polytropic EoS with $n=0.5,1,2$\,\footnote{Note that, at variance with
  Paper~I, we use the polytropic EoS of the form $P=K\rho_0^{1+1/n}$, $\rho=\rho_0+nP$}, and the APR4
EoS~\cite{Akmal:1998cf}. The results of the computation (multiplied by $C^6$) are shown in Fig.~\ref{1love}. We stress
that the computations presented here correct some errors in the numerical implementation of Paper~I. 

\subsection{The hidden symmetry}
Equations~\eqref{eq:prop2}, \eqref{eq:prop2b} imply that the ratios of the RTLNs ${\tilde\lambda}^{(ll')}_E$,
${\tilde\lambda}^{(ll')}_M$ with $l,l'=2,3 $ and $l,l'=3,2$, are constant
\begin{align}
  {\cal R}_1&\equiv\frac{{\tilde\lambda}^{(32)}_M}{{\tilde\lambda}^{(23)}_E}=\frac{9}{4}\sqrt{\frac{5}{7}}\simeq1.902\,,\\
  {\cal R}_2&\equiv\frac{{\tilde\lambda}^{(23)}_M}{{\tilde\lambda}^{(32)}_E}=
  \frac{1}{3}\sqrt{\frac{7}{5}}\simeq0.394\,,\label{eq:prop2r}
\end{align}
regardless of the NS mass and of the EoS. In Fig.~\eqref{fig:ratios} we show the ratios ${\cal R}_1$, ${\cal R}_2$ as
functions of the compactness, for the sample of EoS considered in this paper (polytropic EoS with $n=0.5,1,2$ and APR4
EoS), together with the theoretical prediction~\eqref{eq:prop2r}.

Note that the theoretical prediction from the Lagrangian PN approach [Eq.~\eqref{eq:prop2r}] is precisely satisfied in
the small-compactness limit. When $C\lesssim0.1$, the ratios ${\cal R}_1$, ${\cal R}_2$ coincide with the theoretical
prediction within $1\%$, whereas for $C\sim0.2-0.3$, the discrepancy increases (approximately) quadratically in $C$ up
to $\sim6\%$. We have verified that this discrepancy is significantly larger than the numerical errors, so it does not
look as a numerical artifact. Moreover, for any compactness the ratios ${\cal R}_1$, ${\cal R}_2$ are independent on the
EoS within $0.2\%$ (i.e., within the numerical error).

\section{Discussion} \label{sec:conclusions}
Our results imply that a hidden symmetry among the static RTLNs with opposite parity actually exists, as suggested by
the PN Lagrangian formulation of Paper I. This symmetry is exaclty satisfied in the small-compactness limit,
and is therefore stronger than other approximately EoS-independent relations between the (spin- and tidal- induced)
multipole moments of a NS~\cite{Yagi:2016bkt}.  For large values of the compactness the hidden symmetry is
weakly violated, but an EoS-independent (within numerical errors) relation is still present.

The very existence of this hidden symmetry is highly nontrivial and deserves further studies. In particular, it is not
clear which is its underlying reason. From a Lagrangian point of view it is natural to expect that opposite sectors are
coupled to each other, since a single interaction term in the form of Eq.~\eqref{LagrangianCoupling} gives rise to related
coupling terms in the field equations for ${\cal A}$ and ${\cal P}$ which are both proportional to the single coupling constant
$\alpha$. On the other hand, justifying the origin of this symmetry from a pertubation-theory point of view is
challenging, since there is {\it a priori} no reason why perturbations belonging to opposite parity sectors should be
related to each other.  This symmetry is somehow reminiscent of the relation between axial and polar perturbations in
Schwarzschild black holes found by Chandrasekhar~\cite{Chandrasekhar:1985kt}, although in this case it involves the
matter sector as well.

Another point that deserves future investigation is the dependence on the compactness. The latter does not enter
directly in the Lagrangian formulation of Paper~II, being encoded in the (R)TLNs. Our results suggest instead that the
prediction from the PN expansion is valid only for low-compactness objects, and it acquires small corrections at large
compactness.

Finally, in agreement with the framework of Paper~II, we have focused on {\it static} perturbations. We anyway expect
that  qualitatively similar (but quantitatively different) relations exist among the
\emph{irrotational} RTLNs. In order to extend our results to the more realistic, irrotational case, one should include
time dependence (and, possibly, nonaxisymmetry) in the tidal field, both in the Lagrangian and in the
perturbation-theory formulations. This interesting problem will be discussed in a future work~\cite{inprep}.

\begin{acknowledgments}
P.P. acknowledges financial support provided under the European Union's H2020 ERC, Starting 
Grant agreement no.~DarkGRA--757480. We also acknowledge support under the MIUR PRIN and FARE programmes (GW-NEXT, 
CUP:~B84I20000100001), from the Amaldi Research Center funded by the MIUR program "Dipartimento di 
Eccellenza" (CUP:~B81I18001170001), and networking support by the COST Action CA16104.
\end{acknowledgments}

\appendix
\section{Explicit form of the coefficients and the source in the RTLN equations} \label{app:RTLNs}
In this appendix we give the explicit form of the coefficients in the definition of the operators ${\cal D}^{{\rm
    pol}\,(l)}$, ${\cal D}^{{\rm ax}\,(l)}$ given in Eqs.~\eqref{eq:dpol}, \eqref{eq:dax}, and of the source terms appearing in
Eqs.~\eqref{eq:polarled}, \eqref{eq:axialled}. We have:
\begin{align}
	C_1^{{\rm pol}\,(l)} &= \frac{e^{\lambda (r)} \left[4 \pi  r^2 \big(P(r)-\rho (r)\big)+1\right]+1}{r}\\
	C_0^{{\rm pol}\,(l)}        &= \frac{4 \pi  e^{\lambda (r)} (P(r)+\rho (r))}{\frac{dP}{d\rho}} \\
	& -\frac{\left(8 \pi  r^2 P(r)+1\right)^2 e^{2 \lambda (r)}}{r^2} \nonumber\\
	& -\frac{e^{\lambda (r)} \left[-4 \pi  r^2 \big(13 P(r)+5 \rho (r)\big)+l ^2+l -2\right]+1}{r^2}\nonumber\\
	C_1^{{\rm ax}\,(l)} &= -4 \pi  r e^{\lambda (r)} (P(r)+\rho (r))\\
	C_0^{{\rm ax}\,(l)}& = -\frac{e^{\lambda (r)} \left[8 \pi  r^2 (P(r)+\rho (r))
            \right]}{r^2}\nonumber \\
	& - \frac{e^{\lambda (r)} \left[l ^2+l -2\right]+2}{r^2}\,,
\end{align}
and
\begin{widetext}
\begin{eqnarray}
 S_+^{{\rm pol}\,(l)} &=& \frac{1}{({l}({l}+1)-2) \sqrt{4 {l} ({l}+2)+3}
   {c_s^2} (r-2 {\cal M})^2}  \left[\kappa\left(2-{l}({l}+1)\right) r^3 (r-2 {\cal M}) (P+\rho )
   ({\tilde{\omega}}+\Omega ) {H_0^{({l})}}\right.\nonumber\\
   &&\left.-{c_s^2}
   \left({H_0^{({l})}} \left(2 r {\cal M} \left(-\kappa  r^2 \rho  \left(\left(5
   {{l}({l}+1)}-22\right) \Omega +({l} (5 {l}+9)-6) {\tilde{\omega}}\right)-2 \left({l}
   \left({{l}({l}+1)}-4\right)-2\right) (\Omega -{\tilde{\omega}})\right.\right.\right.\right.\nonumber\\
   &&\left.\left.\left.\left.+\kappa  r^2 P
   \left(\Omega  \left(3 {l} ({l}+5)+32 \kappa  r^2 \rho +14\right)+(14-{l} (13 {l}+25))
   {\tilde{\omega}}\right)+24 \kappa ^2 r^4 \Omega  P^2+8 \kappa ^2 r^4 \Omega 
   \rho ^2\right)\right.\right.\right.\nonumber\\
   &&\left.\left.\left.+r^2 \left(\kappa  r^2 \Omega  \left(\rho  \left(5 {{l}({l}+1)}+16
   \kappa  r^2 P \left(\kappa  r^2 P-1\right)-14\right)+P \left(4 \kappa  r^2
   P \left({l} ({l}+3)+4 \kappa  r^2 P-2\right)+({l}-3) {l}-6\right)\right.\right.\right.\right.\right.\nonumber\\
   &&\left.\left.\left.\left.\left.-8 \kappa  r^2 \rho
   ^2\right)+{\tilde{\omega}} \left(\kappa  r^2 \left(P \left(-4 \kappa  {l}
   ({l}+3) r^2 P+{l} (9 {l}+13)-14\right)+({l} (5 {l}+9)-6) \rho \right)-2 {l}
   ({l}({l}+1)-2)\right)\right.\right.\right.\right.\nonumber\\
   &&\left.\left.\left.\left. +2 {l} ({l}({l}+1)-2) \Omega \right)+4 {\cal M}^2
   \left((({l}-1) {l}-4) (\Omega -{\tilde{\omega}})-4 \kappa  r^2 \Omega  (P+\rho
   )\right)\right)\right.\right.\nonumber\\
   &&\left.\left.+2 r (r-2 {\cal M}) {H_0^{({l})}}' \left(({l}({l}+1)-2) r
   ({\tilde{\omega}}-\Omega )+{\cal M} \left(({l} (3 {l}+5)-4) (\Omega -{\tilde{\omega}})+4 \kappa  r^2 \Omega
   (P+\rho )\right)\right.\right.\right.\nonumber\\
   &&\left.\left.\left.+\kappa  r^3 P \left({l} ({l}+3)
   (\Omega -{\tilde{\omega}})+4 \kappa  r^2 \Omega  \rho \right)+4 \kappa ^2
   r^5 \Omega  P^2\right)\right)\right]\,,
   \end{eqnarray}
   \begin{eqnarray}
     S_-^{{\rm pol}\,({l})} &=& \frac{1}{({l}({l}+1)-2) \sqrt{4 {l}^2-1} {c_s^2} (r-2 {\cal M})^2}\left[ {c_s^2}
       \left(2 r (r-2 {\cal M}) {H_0^{({l})}}' \left({\cal M} \left(\left(3
   {{l}({l}+1)}-6\right) (\Omega -{\tilde{\omega}})+4 \kappa  r^2 \Omega (P+\rho)\right)\right.\right.\right.\nonumber\\
   &&\left.\left.\left.+r \left(\kappa  r^2 P \left(\left({{l}({l}-1)}-2\right)
   (\Omega -{\tilde{\omega}})+4 \kappa  r^2 \Omega  \rho
   \right)-({l}({l}+1)-2) (\Omega -{\tilde{\omega}})+4 \kappa ^2 r^4
   \Omega  P^2\right)\right) \right.\right.\nonumber\\
   &&\left.\left.+{H_0^{({l})}} \left(4 {\cal M}^2 \left(\left({l}^2+3
   {l}-2\right) (\Omega -{\tilde{\omega}})-4 \kappa  r^2 \Omega  P-4 \kappa  r^2
   \Omega  \rho \right)+2 r {\cal M} \left(\kappa  r^2 P \left(-\left(13
   {{l}({l}+1)}-26\right) {\tilde{\omega}} \right.\right.\right.\right.\right.\nonumber\\
   &&\left.\left.\left.\left.\left. +\left(3 {l}^2-9 {l}+2\right) \Omega +32 \kappa 
   r^2 \Omega  \rho \right)-\kappa  r^2 \rho  \left(\left(5 {{l}({l}+1)}-10\right)
   {\tilde{\omega}}+\left(5 {l}^2+9 {l}-18\right) \Omega \right) \right.\right.\right.\right.\nonumber\\
   &&\left.\left.\left.\left. +2 \left({l}^3+2 {l}^2-3
   {l}-2\right) (\Omega -{\tilde{\omega}})+24 \kappa ^2 r^4 \Omega  P^2+8 \kappa
   ^2 r^4 \Omega  \rho ^2\right)+r^2 \left(\kappa  r^2 P \left(\left(9 {l}^2+5
   {l}-18\right) {\tilde{\omega}}\right.\right.\right.\right.\right.\nonumber\\
   &&\left.\left.\left.\left.\left. +\left({l}^2+5 {l}-2\right) \Omega -16 \kappa  r^2
   \Omega  \rho \right)+4 \kappa ^2 r^4 P^2 \left(\left(-{{l}({l}+1)}+2\right)
   {\tilde{\omega}}+\left({{l}({l}-1)}-4\right) \Omega +4 \kappa  r^2 \Omega  \rho
   \right)\right.\right.\right.\right. \nonumber\\
   &&\left.\left.\left.\left. +\kappa  r^2 \rho  \left(\left(5 {{l}({l}+1)}-10\right) {\tilde{\omega}}+
   \left(5 {l}^2+9 {l}-10\right) \Omega \right)-2 \left({l}^3+2 {{l}({l}-1)}-2\right)
   (\Omega -{\tilde{\omega}})+16 \kappa ^3 r^6 \Omega  P^3-8 \kappa ^2 r^4
   \Omega  \rho ^2\right)\right)\right)\right.\nonumber\\
   &&\left.+\kappa  ({l}({l}+1)-2) r^3 (r-2 {\cal M})    (P+\rho ) ({\tilde{\omega}}+\Omega )
   {H_0^{({l})}}\right]\,,
      \end{eqnarray}
   \begin{eqnarray}
     S_+^{{\rm ax}\,({l})} &=&  \frac{2 e^{-\nu }}{\sqrt{4 {l}^2+8 {l}+3} r^2 {c_s^2} (r-2{\cal M})^2} \left[{c_s^2}
       \left(r (r-2 {\cal M}) {h_0^{({l})}}'
   \left(-2 {\cal M} \left(r {\tilde{\omega}}' \left({{l}({l}+1)}+2 \kappa  r^2 P+2 \kappa
    r^2 \rho +3\right)\right.\right.\right.\right.\nonumber\\
    &&\left.\left.\left.\left.+{\tilde{\omega}} \left(3 {l}({l}+1)-4 \kappa  r^2 P-4
   \kappa  r^2 \rho \right)-3 {l}({l}+1) \Omega \right)+r \left(2 \kappa  r^2 P
   \left({l}({l}+1) \Omega -{\tilde{\omega}} \left({{l}({l}+1)}-4 \kappa  r^2 \rho
   \right)\right)+{l}^2 r {\tilde{\omega}}'\right.\right.\right.\right.\nonumber\\
   &&\left.\left.\left.\left.-2 {l}^2 \Omega +{l} r {\tilde{\omega}}'+2 {l}({l}+1) {\tilde{\omega}}-2 {l}
   \Omega -2 \kappa ^2 r^4 P^2 \left(r
   {\tilde{\omega}}'-4 {\tilde{\omega}}\right)+2 \kappa  r^3 \rho 
   {\tilde{\omega}}'\right)+10 {\cal M}^2 {\tilde{\omega}}'\right)\right.\right.\nonumber\\
   &&\left.\left.+{h_0^{({l})}} \left(r^2 \left(\kappa  r^2 P \left(2
   {\tilde{\omega}} \left({{l}({l}+1)}+12 \kappa  r^2 \rho \right)-2 {\tilde{\omega}}' \left({l} r-4
   \kappa  r^3 \rho +r\right)-5 {l}({l}+1) \Omega \right)+\kappa 
   r^2 \rho  \left(2 {l}({l}+1) {\tilde{\omega}}\right.\right.\right.\right.\right.\nonumber\\
   &&\left.\left.\left.\left.\left.-5 {l}({l}+1) \Omega -4 r
   {\tilde{\omega}}'\right)+2 {l}({l}+1)^2 (\Omega -{\tilde{\omega}})-20 \kappa
   ^3 r^6 P^3 {\tilde{\omega}}-4 \kappa ^2 r^4 P^2 \left({\tilde{\omega}} \left(5 \kappa  r^2 \rho -4\right)-
   3 r {\tilde{\omega}}'\right)+8
   \kappa ^2 r^4 \rho ^2 {\tilde{\omega}}\right)\right.\right.\right.\nonumber\\
   &&\left.\left.\left.-2 r {\cal M} \left(-2 {l}^3
   {\tilde{\omega}}+2 {l}^3 \Omega +\kappa  r^2 P \left(2 {\tilde{\omega}}
   \left({{l}({l}+1)}+22 \kappa  r^2 \rho +2\right)+2 r {\tilde{\omega}}' \left(4 \kappa  r^2 \rho -5-{l}\right)
   -5 {l}({l}+1) \Omega \right)\right.\right.\right.\right.\nonumber\\
   &&\left.\left.\left.\left.+\kappa  r^2 \rho  \left(2
   \left({{l}({l}+1)}+2\right) {\tilde{\omega}}-5 {l}({l}+1) \Omega -12 r {\tilde{\omega}}'\right)
   -2 {l}^2 {\tilde{\omega}}+2 {l}^2 \Omega +{l} r {\tilde{\omega}}'+12 \kappa ^2 r^4 P^2
   \left(r {\tilde{\omega}}'+3 {\tilde{\omega}}\right)\right.\right.\right.\right.\nonumber\\
   &&\left.\left.\left.\left.+8 \kappa ^2 r^4 \rho ^2 {\tilde{\omega}}-5 r {\tilde{\omega}}'\right)
   -4 {\cal M}^2 \left(r {\tilde{\omega}}' \left(-{l}+8 \kappa  r^2 P+8
   \kappa  r^2 \rho +10\right)+{\tilde{\omega}} \left(-2 {l}({l}+1)+\kappa  r^2
   P+\kappa  r^2 \rho \right)+2 {l}({l}+1) \Omega \right)\right.\right.\right.\nonumber\\
   &&\left.\left.\left.+40 {\cal M}^3 {\tilde{\omega}}'\right)\right)-\kappa  r^2 (P+\rho ) {h_0^{({l})}} \left(-2 r {\cal M}
   \left({l}({l}+1) \Omega -4 \kappa  r^2 P {\tilde{\omega}}\right)+r^2 \left({l}
   ({l}+1) \Omega +4 \kappa ^2 r^4 P^2 {\tilde{\omega}}\right)+4 {\cal M}^2
   {\tilde{\omega}}\right)\right]\,,
      \end{eqnarray}
   \begin{eqnarray}
     S_-^{{\rm ax}\,({l})} &=& \frac{2 e^{-\nu }}{\sqrt{4 {l}^2-1} r^2 {c_s^2} (r-2 {\cal M})^2} \left[{c_s^2}
       \left({h_0^{({l})}} \left(r^2
   \left(-\kappa  r^2 P \left(2 {\tilde{\omega}} \left({{l}({l}+1)}+12 \kappa  r^2
   \rho \right)+2 r {\tilde{\omega}}' \left({l}+4 \kappa  r^2 \rho \right)-5 {l}
   ({l}+1) \Omega \right)\right.\right.\right.\right.\nonumber\\
   &&\left.\left.\left.\left.+2 {l}^2 ({l}+1) (\Omega -{\tilde{\omega}})+\kappa  r^2 \rho
   \left(-2 {l}({l}+1) {\tilde{\omega}}+5 {l}({l}+1) \Omega +4 r {\tilde{\omega}}'\right)+20 \kappa ^3
   r^6 P^3 {\tilde{\omega}}\right.\right.\right.\right. \nonumber\\
    &&\left.\left.\left.\left.+4 \kappa ^2 r^4 P^2
   \left({\tilde{\omega}} \left(5 \kappa  r^2 \rho -4\right)-3 r {\tilde{\omega}}'\right)-8 \kappa ^2
   r^4 \rho ^2 {\tilde{\omega}}\right)+2 r {\cal M}
   \left(2 {l}^3 {\tilde{\omega}}-2 {l}^3 \Omega +\kappa  r^2 P \left(2
   {\tilde{\omega}} \left({{l}({l}+1)}+22 \kappa  r^2 \rho +2\right)\right.\right.\right.\right.\right.\nonumber\\
   &&\left.\left.\left.\left.\left.+2 r {\tilde{\omega}}' \left({l}+4 \kappa  r^2 \rho -4\right)-
   5 {l}({l}+1) \Omega \right)+\kappa 
   r^2 \rho  \left(2 \left({{l}({l}+1)}+2\right) {\tilde{\omega}}-5 {l}({l}+1) \Omega -12
   r {\tilde{\omega}}'\right)+4 {l}^2 (\tilde{\omega}-\Omega) \right.\right.\right.\right. \nonumber\\
   &&\left.\left.\left.\left.-{l} r
   {\tilde{\omega}}'+2 {l} {\tilde{\omega}}-2 {l} \Omega +12 \kappa ^2 r^4 P^2
   \left(r {\tilde{\omega}}'+3 {\tilde{\omega}}\right)+8 \kappa ^2 r^4 \rho
   ^2 {\tilde{\omega}}-6 r {\tilde{\omega}}'\right)+4 {\cal M}^2 \left(r
   {\tilde{\omega}}' \left({l}+8 \kappa  r^2 P+8 \kappa  r^2 \rho
   +11\right)\right.\right.\right.\right.\nonumber\\
   &&\left.\left.\left.\left.+{\tilde{\omega}} \left(-2 {l}({l}+1)+\kappa  r^2 P+\kappa  r^2
   \rho \right)+2 {l}({l}+1) \Omega \right)-40 {\cal M}^3 {\tilde{\omega}}'\right)-r
   (r-2 {\cal M}) {h_0^{({l})}}' \left(-2 {\cal M} \left(r {\tilde{\omega}}'
   \left({{l}({l}+1)}\right.\right.\right.\right.\right.\nonumber\\
   &&\left.\left.\left.\left.\left.+2 \kappa  r^2 (P+\rho) +3\right)+{\tilde{\omega}}
   \left(3 {l}({l}+1)-4 \kappa  r^2 P-4 \kappa  r^2 \rho \right)-3 {l}({l}+1) \Omega
   \right)\right.\right.\right.\nonumber\\
   &&\left.\left.\left.+r \left(2 \kappa  r^2 P \left({l}({l}+1) \Omega -{\tilde{\omega}}
   \left({{l}({l}+1)}-4 \kappa  r^2 \rho \right)\right)+{l}^2 r {\tilde{\omega}}'-2 {l}^2
   \Omega +{l} r {\tilde{\omega}}'+2 {l}({l}+1) {\tilde{\omega}}-2 {l} \Omega\right.\right.\right.\right.\nonumber\\
   &&\left.\left.\left.\left.-2
   \kappa ^2 r^4 P^2 \left(r {\tilde{\omega}}'-4 {\tilde{\omega}}\right)+2
   \kappa  r^3 \rho  {\tilde{\omega}}'\right)+10 {\cal M}^2 \tilde{\omega}'\right)\right)+
   \kappa  r^2 (P+\rho ) {h_0^{({l})}} \left(-2 r {\cal M}
   \left({l}({l}+1) \Omega -4 \kappa  r^2 P {\tilde{\omega}}\right)\right.\right.\nonumber\\
   &&\left.\left.+r^2 \left({l}
   ({l}+1) \Omega +4 \kappa ^2 r^4 P^2 {\tilde{\omega}}\right)+4 {\cal M}^2
   {\tilde{\omega}}\right)\right]\,,
\end{eqnarray}
\end{widetext}
where $\kappa=4\pi$ and $c_s=\sqrt{dP/d\rho}$ is the sound speed in the fluid.

\section{Comparison between different conventions for the (R)TLNs}\label{app:conversion}
We shall compare the definitions and notations of TLNs and RTLNs in Paper~I and in Paper~II, in order to find the
rescaling factors appearing in Eq.~\eqref{eq:transf}. For the reader's convenience, we repeat here some of the relations
that appear in the main text, so that the derivation is self-contained.

In Paper~I the adiabatic relations among multipole moments read [see Eq.~\eqref{eq:adiabrelp}], 
\begin{align}
  \frac{M_2}{M^3}&=\tilde{\lambda}^{(2)}_E {\cal E}^{(2)}_0M^2+\chi\tilde\lambda^{(23)}_E{\cal B}_0^{(3)}M^3\nonumber\\
  \frac{M_3}{M^4}&=\tilde{\lambda}^{(3)}_E{\cal E}^{(3)}_0M^3+\chi\tilde\lambda^{(32)}_E{\cal B}_0^{(2)}M^2\nonumber\\
  \frac{J_2}{M^3}&=\tilde{\lambda}^{(2)}_M{\cal B}^{(2)}_0M^2+\chi\tilde\lambda^{(23)}_M{\cal E}_0^{(3)}M^3\nonumber\\
  \frac{J_3}{M^4}&=\tilde{\lambda}^{(3)}_M{\cal B}^{(3)}_0M^3+\chi\tilde\lambda^{(32)}_M{\cal E}_0^{(2)}M^2\,,
  \label{eq:adiabrelp2}
\end{align}
where $M_l$, $J_l$ are the Geroch-Hansen multipole moments and ${\cal E}^{(l)}_0$, ${\cal B}_0^{(l)}$ are the tidal
tensor components defined from the asymptotic limit of the metric in Eq.~\eqref{eq:defcEcB}. Conversely, in Paper~II they
read~\eqref{eq:adiabatic} (in $c=1$ units)
\begin{align}
Q^{ab}&=\lambda_2 G^{ab}+\lambda_{23}J^cH^{abc}\,,\nonumber\\
Q^{abc}&=\lambda_3 G^{abc}+\lambda_{32}J^{<c} H^{ab>}\,,\nonumber\\
S^{ab}&=\sigma_2 H^{ab}+\sigma_{23}J^cG^{abc}\,,\nonumber\\
S^{abc}&=\sigma_3 H^{abc}+\sigma_{32}J^{<c} G^{ab>}\,,\label{eq:adiabatic2}
\end{align}
where $Q^L$, $S^L$ are Thorne's multipole moments, and $G^L$, $H^L$ are the tidal tensor components, defined (in
spherical harmonic notation) from the asymptotic limit expansion~\eqref{eq:defcEcB2}. In order to express
Eqs.~\eqref{eq:adiabatic2} in terms of spherical harmonics components, we note that (see Sec.~\eqref{subsec:notation})
\begin{align}
  Q^{ab}&=\sum_mQ^{2m}{\cal Y}^{2 m}_{ab}=Q^{20}{\cal Y}^{20}_{ab}  \nonumber\\
  Q^{abc}&=\sum_mQ^{2m}{\cal Y}^{3 m}_{abc}=Q^{30}{\cal Y}^{30}_{abc}
\end{align}
where we have assumed axisymmetry; the same applies to the current multipole moments $S^{ab}$, $S^{abc}$ and to the
tidal tensor components. Moreover, $J^a=\chi k^aM^2$. Therefore,
\begin{align}
Q^{20}{\cal Y}^{20}_{ab}&=\lambda_2 G^{20}{\cal Y}^{20}_{ab}+\lambda_{23}\chi M^2H^{30}{\cal Y}^{30}_{abc}k^c\nonumber\\
Q^{30}{\cal Y}^{30}_{abc}&=\lambda_3 G^{30}{\cal Y}^{30}_{abc}+\lambda_{32}\chi M^2H^{20}{\cal Y}^{20}_{<ab}k_{c>}\nonumber\\
S^{20}{\cal Y}^{20}_{ab}&=\sigma_2 H^{20}{\cal Y}^{20}_{ab}+\sigma_{23}\chi M^2G^{30}{\cal Y}^{30}_{abc}k^c\nonumber\\
S^{30}{\cal Y}^{30}_{abc}&=\sigma_3 H^{30}{\cal Y}^{30}_{abc}+\sigma_{32}\chi M^2G^{20}{\cal Y}^{20}_{<ab}k_{c>}\,.
\end{align}
Since ${\cal Y}^{lm}_{ab}{\cal Y}^{*lm'}_{ab}=N_l^{-1}\delta^{mm'}$ with $N_l=4\pi l!/(2l+1)!!$, we find
\begin{align}
Q^{20}&=\lambda_2 G^{20}+\lambda_{23}\chi M^2H^{30}KN_2\nonumber\\
Q^{30}&=\lambda_3 G^{30}+\lambda_{32}\chi M^2H^{20}KN_3  \nonumber\\
S^{20}&=\sigma_2 H^{20}+\sigma_{23}\chi M^2G^{30}KN_2\nonumber\\
S^{30}&=\sigma_3 H^{30}+\sigma_{32}\chi M^2G^{20}KN_3\,.\label{eq:adiab3}
\end{align}
where we defined $K={\cal Y}^{30}_{abc}k^c{\cal Y}^{*20}_{ab}$. Since\,\cite{PoissonWill}
${\cal Y}^{20}_{ab}=-\sqrt{5/(16\pi)}\delta_{ab}$ and ${\cal Y}^{30}_{abc}k^c=\sqrt{7/(16\pi)}\delta_{ab}$,
\begin{equation}\label{eq:defK}
K=\frac{3}{16\pi}\sqrt{35}\,.
\end{equation}
The Thorne multipole moments are related to the Geroch-Hansen multipole moments by Eq.~\eqref{eq:transfmult},
\begin{equation}
  Q^{i_1\cdots i_l}=M_lk^{i_1\cdots i_l}~~,~~~~~
  S^{i_1\cdots i_l}=\frac{l+1}{2l}J_lk^{i_1\cdots i_l}\,,
\end{equation}
thus, since (see Sec.~\eqref{subsec:notation})
$Q^{i_1\cdots i_l}n_{i_1\cdots i_l}=\sum_mQ^{lm}Y^{lm}=Q^{l0}Y^{l0}$ and the same applies to $S^{i_1\cdots i_l}$,
\begin{align}
  M_lk^{i_1\cdots i_l}n_{i_1\cdots i_l}&=\sqrt{\frac{2l+1}{4\pi}}Q^{l0}P_l(\cos\theta)\nonumber\\
  \frac{l+1}{2l}J_lk^{i_1\cdots i_l}n_{i_1\cdots i_l}&=\sqrt{\frac{2l+1}{4\pi}}S^{l0}P_l(\cos\theta)\,,
\end{align}
where $P_l$ are the Legendre polynomials. Then, since\,\cite{PoissonWill}
$k^{i_1\cdots i_l}n_{i_1\cdots i_l}=l!/(2l-1)!!P_l(\cos\theta)$,
\begin{align}
  Q^{l0} &= \frac{l!}{(2l-1)!!}\sqrt{\frac{4\pi}{2l+1}}M_l\nonumber\\
  S^{l0} &= \frac{(l+1)!}{2l(2l-1)!!}\sqrt{\frac{4\pi}{2l+1}} J_l\,.\label{eq:transfQSMJ}
\end{align}
Moreover, by comparing Eqs.~\eqref{eq:defcEcB}, \eqref{eq:defcEcB2} we find that the tidal tensor components in the
notations of Paper~I and of Paper~II are related by
\begin{align}
  G^{lm} &=  -(l-2)!{\cal E}_m^{(l)}\nonumber\\
  H^{lm} &= -\frac{2}{3}(l+1)(l-2)!{\cal B}_m^{(l)}\,.\label{eq:tidalconversion}
\end{align}
By replacing Eqs.~\eqref{eq:transfQSMJ}, \eqref{eq:tidalconversion} in Eq.~\eqref{eq:adiab3} we find
\begin{align}
  M_2 &= -\frac{3}{2}\sqrt{\frac{5}{4\pi}}\lambda_2{\cal E}_0^{(2)} -
  \frac{32\pi}{15}\sqrt{\frac{5}{4\pi}}\chi M^2K\lambda_{23}{\cal B}_0^{(3)}\nonumber\\
  M_3 &= -\frac{5}{2}\sqrt{\frac{7}{4\pi}}\lambda_3{\cal E}_0^{(3)} -
  \frac{8\pi}{7}\sqrt{\frac{7}{4\pi}}\chi M^2K\lambda_{32}{\cal B}_0^{(2)}\nonumber\\
  J_2 &= -4\sqrt{\frac{5}{4\pi}}\sigma_2{\cal B}_0^{(2)} -
  \frac{16\pi}{15}\sqrt{\frac{5}{4\pi}}\chi M^2K\sigma_{23}{\cal E}_0^{(3)}\nonumber\\
  J_3 &= -10\sqrt{\frac{7}{4\pi}}\sigma_3{\cal B}_0^{(3)} -
  \frac{6\pi}{7}\sqrt{\frac{5}{4\pi}}\chi M^2K\sigma_{32}{\cal E}_0^{(2)}\,.
\end{align}
By comparing with Eq.~\eqref{eq:adiabrelp2} and replacing Eq.~\eqref{eq:defK}, we finally obtain Eqs.~\eqref{eq:transf},
i.e. the relation between TLNs and RTLNs in the notations of Paper~I and Paper~II.

\bibliographystyle{apsrev4-1}
\bibliography{biblio}

\end{document}